\documentclass{emulateapj}

\usepackage{subequation}
\journalinfo{ApJ, in press (v. 661, June 2007)}

\newcommand{\frequnits}{~$\rm s^{-1}$}
\newcommand{\angfrequnits}{~$\rm rad~s^{-1}$}
\newcommand{\lengthunits}{~$R_\odot$}
\newcommand{\velunits}{~$\rm km~s^{-1}$}
\newcommand{\elsasser}{Els\"{a}sser}
\def\vec#1{\ensuremath{\mathbf{#1}}}

\shorttitle{Alfv\'en Waves in Multicomponent Winds}
\shortauthors{Li and Li}

\begin{document}
\title{Propagation of non-WKB Alfv\'en waves in a multicomponent solar wind with differential ion flow}
\author{Bo Li and Xing Li}
\affil{Institute of Mathematical and Physical Sciences, University of Wales Aberystwyth,
  SY23 3BZ, UK}
\email{bbl@aber.ac.uk}

\begin{abstract}
The propagation of dissipationless, hydromagnetic,
      purely toroidal Alfv\'en waves
      in a realistic background three-fluid solar wind with axial symmetry and
      differential proton-alpha flow is investigated.
The short wavelength WKB approximation is not invoked.
Instead, the equations that govern the wave transport are derived from standard
      multi-fluid equations in the five-moment approximation. 
The Alfv\'enic point, where the combined poloidal Alfv\'en Mach number $M_T=1$,
      is found to be a singular point for the wave equation,
      which is then numerically solved
      for three representative angular frequencies
      $\omega=10^{-3}$, $10^{-4}$ and $10^{-5}$~rad~s$^{-1}$
      with a fixed wave amplitude of 10~km~s$^{-1}$
      imposed at the coronal base (1~$R_\odot$).
The wave energy and energy flux densities as well as wave-induced ion acceleration
      are computed and compared with those derived in the WKB limit.
Between 1~$R_\odot$ and 1~AU, the
      numerical solutions show substantial deviation from the WKB expectations.
Even for the relatively high frequency $\omega=10^{-3}$\angfrequnits,
      a WKB-like behavior can be seen only in regions $r\gtrsim 10$~$R_\odot$.
In the low-frequency case $\omega=10^{-5}$\angfrequnits, the computed profiles
      of wave-related parameters show a spatial dependence distinct from the WKB one, 
      the deviation being particularly pronounced
      in interplanetary space.
In the inner corona $r\lesssim 4$~$R_\odot$, 
      the computed ion velocity fluctuations are
      considerably smaller than the WKB expectations in all cases,
      as is the computed wave-induced acceleration exerted on 
      protons or alpha particles.
As for the wave energy and energy flux densities, they can be enhanced
      or depleted compared
      with the WKB results, depending on $\omega$.
With the chosen base wave amplitude,
      the wave acceleration has negligible effect on the ion force balance
      in the corona.
Hence processes other than the non-WKB wave acceleration are needed to accelerate
      the ions out of the gravitational potential well of the Sun.
However, at large distances beyond the Alfv\'enic point, the low-frequency waves can play an important role
      in the ion dynamics, with the net effect being to equalize the speeds of 
      the two ion species considered.
\end{abstract}

\section{INTRODUCTION}
Ever since their identification by \citet{BelcherDavis_71}, 
     Alfv\'en waves have been extensively studied using {\it in situ} measurements,
     such as by Helios and Ulysses, covering
     the heliocentric distance from 0.29 to 4.3~AU 
     \citep{TuMarsch_95, Goldstein_etal_95, Bavassano_etal_00a, Bavassano_etal_00b}.
On the other hand, the non-thermal broadening of a number of Ultraviolet lines,
     such as those measured with 
     the SUMER (Solar Ultraviolet Measurements of Emitted Radiation)
     and UVCS (Ultraviolet Coronagraph Spectrometer) instruments
     on SOHO (the Solar and Heliospheric Observatory),
     is usually attributed to the transverse velocity fluctuations, thereby enabling one
     to infer the amplitudes of these fluctuations in the 
     inner corona below $\sim 5$\lengthunits\ \citep{Banerjee_etal_98, Esser_etal_99}.
Moreover, the Faraday rotation measurements, which yield information regarding 
     the line-of-sight magnetic field fluctuations, have been shown to support indirectly 
     the presence of Alfv\'en waves inside 10~$R_\odot$ \citep{Hollweg_etal_82}.
The solar wind in intermediate regions, for the time being, can be explored only by radio scintillation
     measurements which allow one to derive the velocity fluctuations
     \citep[e.g.,][]{ArmstrongWoo_81, Scott_etal_83}.
It is noteworthy that although the hourly-scale fluctuations seem to be more 
     frequently studied, the fluctuation spectrum measured by Helios
     nevertheless spans a broad frequency range from $10^{-5}$ to $10^{-2}$\frequnits\
     \citep{TuMarsch_95}.

Most of the theoretical investigations into the interaction between
      Alfv\'en waves and the solar wind have been performed in
      the short wavelength WKB limit which makes the problem
      more tractable mathematically.
For instance, by employing the WKB approximation, \citet{Parker_65} derived 
      an expression for the ponderomotive force through which
      the Alfv\'en waves may provide further acceleration to 
      the solar wind.
The wave acceleration was later incorporated in detailed
      numerical models by, e.g., \citet{AlazrakiCouturier_71}.
It was soon realized that Alfv\'en waves may also heat
      the solar wind via dissipative processes such as the cyclotron resonance interaction
      between ions and high frequency, parallel propagating waves generated by
      a turbulent cascade (cf. the extensive review by \citet{HI02}).
Such a parallel cascade scenario has been successful in explaining a number
      of observations, to name but one, 
      the significant thermal anisotropy of ions as established by UVCS measurements
      \citep{Li_etal_99}.
As pointed out by \citet{HI02}, the applicability of the WKB approximation
      in which the processes are formulated
      is questionable in the near-Sun region in view of the large Alfv\'en speeds.
Furthermore, a turbulent cascade requires a non-vanishing magnetic Reynolds
      stress tensor, which, however, is zero in the WKB limit since
      the particle and field components of the tensor cancel each other exactly.
A non-WKB analysis is therefore required to account for the wave reflection 
      and the consequent driving of any turbulence cascade.

As a matter of fact, non-WKB analysis of Alfv\'en waves in the solar wind has been carried out
      for decades \citep[e.g.,][]{HO80, Lou_93}.
This, however, is almost exclusively done in the framework of ideal MHD, which allows 
      waves propagating in opposite directions to be explicitly separated when 
      the \elsasser\ variables are used. 
The adoption of the \elsasser\ variables has also enabled a new turbulence 
      phenomenology concerning the nonlinear coupling between counter-propagating waves
      \citep{Dmitruk_etal_01, CranmerBalle_05, Verdini_etal_05}.
This coupling term, if interpreted as the energy cascaded towards fluctuations with 
      increasingly large perpendicular wavenumbers, is also more consistent with
      theoretical expectations. 

Despite substantial advances achieved in ideal MHD, which is appropriate
      for the description of the gross properties of the solar wind,
      the non-WKB analysis of Alfv\'en waves has rarely been done
      using multi-fluid transport equations.
In contrast, multi-fluid, Alfv\'en wave driven solar wind modeling
      formulated in the WKB limit
      has reached considerable sophistication \citep[see][]{HI02}. 
Such a multi-fluid approach is particularly necessary for the solar wind
      since the alpha particles
      must be included given
      their non-negligible abundance and the fact that the
      proton-alpha differential speed can be a substantial fraction of 
      the proton speed in the fast stream
      \citep{Marsch_etal_82}.
There is therefore an obvious need to extend available non-WKB analyses
      of Alfv\'en waves
      from the ideal MHD to the multi-fluid case.

The intent of this paper is to present an analysis of Alfv\'en waves
      in a 3-fluid solar wind assuming axial symmetry, without
      assuming that the wavelength is small compared with the spatial
      scales at which the background parameters vary.
The perturbed velocity and magnetic field are assumed to be
      in the azimuthal direction, i.e, 
      only purely toroidal waves are investigated.
To further simplify the treatment, the wave dissipation is neglected.
This simplification is necessary here since if one wants to gain some
      quantitative insights into the wave dissipation, and 
      to maintain a reasonable self-consistency at the same time,
      one should formulate the dissipation in terms of the amplitudes
      of waves propagating outwards and inwards.
To this end, a full multi-fluid \elsasser\ analysis is required but is unfortunately
      unavailable at the present time.
However, if the wave dissipation is neglected and therefore
      the nonlinear interaction between waves and the multicomponent wind
      is entirely through the agent of ponderomotive forces,
      then the problem can be formulated without distinguishing 
      explicitly between the directions of wave propagation \citep[cf.][]{Lou_93}.

The paper is organized as follows.
In section~\ref{sec_mathform}, we show how to reduce 
      the general multi-fluid transport equations to the desired form.
The resultant equations governing the Fourier amplitudes
      of toroidal Alfv\'en waves at a given frequency
      are then solved analytically in two limiting cases, namely the 
      WKB and zero-frequency limits, in section~\ref{sec_ana_sol}.
Apart from these analytically tractable cases, the equations have to be
      solved numerically.
In section~\ref{sec_num_model}, we re-formulate the model equations
      for numerical convenience, describe the background flow parameters
      and detail the solution procedure as well.
The numerical solutions for three different frequencies are presented 
      in section~\ref{sec_num_res}.
Finally, section~\ref{sec_conc} summarizes the results, ending with some
      concluding remarks.

\section{MATHEMATICAL FORMULATION}
\label{sec_mathform}

Presented in this section is the mathematical development of
     the equations that 
     govern the toroidal fluctuations in a solar wind
     which consists of electrons ($e$), protons
     ($p$) and alpha particles ($\alpha$).
Each species $s$ ($s = e, p, \alpha$) is characterized by its mass $m_s$,
     electric charge $e_s$, number density
     $n_s$, mass density $\rho_s = n_s m_s$, velocity $\vec{v}_s$,
     and partial pressure $p_s$.
If measured in units of the electron charge $e$, $e_s$ may be expressed by
     $e_s = Z_s e$ with $Z_e \equiv -1$ by definition.

To simplify the mathematical treatment, a number of assumptions have been made
     and are collected as follows:
\begin{enumerate}
\item
It is assumed that the solar wind can be described by the standard transport equations
     in the five-moment approximation.
\item 
Quasi neutrality and quasi-zero current are assumed, 
      i.e., $n_e = \sum_k Z_k n_k$
      and $\vec{v}_e = \sum_k Z_k n_k \vec{v}_k/n_e$ where $k=p, \alpha$.
\item 
Symmetry about the magnetic axis is assumed, i.e.,
     $\partial/\partial\phi\equiv 0$ in a heliocentric spherical coordinate system
     ($r, \theta, \phi$).
\item 
The time-independent solar wind interacts with the waves only through
     the wave-induced ponderomotive forces.
\item 
The wave frequency considered is in the hydromagnetic regime, i.e., well below
    the ion gyro-frequencies.
\item 
The perturbed velocity and magnetic field are assumed to be in the $\phi$ direction only.
\item
The effects of the solar rotation on the background solar wind are neglected such that
     there is no need to consider the coupling of Alfv\'en waves to the compressional modes 
     in the presence of a spiral magnetic field.
\item 
The effects of the Coulomb friction on the waves are neglected, 
     and so is the wave-induced modification of the Coulomb friction between background ion flows.
\end{enumerate}

\subsection{Multi-fluid Equations}
\label{sec_geneq}

The equations appropriate for a multi-component solar wind
    plasma in the standard five-moment approximation
    are as follows
    (for the derivation see appendix A.1 in \citet{LiLi06})
\begin{eqnarray}
    \frac{\partial n_k}{\partial t} 
&+& \nabla\cdot(n_k \vec{v}_k) = 0,  \label{gen_nk}  \\
    \frac{\partial \vec{v}_k}{\partial t}
&+& \vec{v}_k\cdot\nabla\vec{v}_k
     +\frac{\nabla p_k}{n_k m_k} +\frac{Z_k\nabla p_e}{n_e m_k} 
     +\frac{GM_\odot}{r^2}\hat{r} \nonumber \\
&-& \frac{1}{n_k m_k}\left[\frac{\delta\vec{M}_k}{\delta t}
     +\frac{Z_k n_k}{n_e}\frac{\delta\vec{M}_e}{\delta t} \right] \nonumber \\
&-& \frac{Z_k}{4\pi n_e m_k}\left(\nabla\times\vec{B}\right)\times\vec{B} \nonumber \\
&+& \frac{Z_k e}{m_k c} \frac{n_j Z_j}{n_e} \left(\vec{v}_j-\vec{v}_k\right)\times\vec{B}   
    =0 ,   \label{gen_vec_vk}  \\
    \frac{\partial }{\partial t} \frac{p_s}{\gamma -1} 
&+& \vec{v}_s\cdot\nabla\frac{p_s}{\gamma -1} 
    +\frac{\gamma}{\gamma-1}p_s(\nabla\cdot\vec{v}_s)  \nonumber \\
&+&  \nabla\cdot\vec{q}_s -\frac{\delta E_s}{\delta t} -Q_s = 0,      \label{gen_ps} \\
     \frac{\partial \vec{B}}{\partial t}
&-&\nabla\times\left(\vec{v}_e \times \vec{B}\right) = 0, \label{gen_vec_magind}
\end{eqnarray}
     where the subscript $s$ refers to all species ($s=e,p,\alpha$), 
     while $k$ stands for ion species only ($k = p, \alpha$).
The gravitational constant is denoted by $G$,
     $M_\odot$ is the mass of the Sun,
     $\vec{B}$ the magnetic field and $c$ the speed of light.
The momentum and energy exchange rates due to the Coulomb
     collisions of species $s$ with the remaining ones
     are denoted by $\delta \vec{M}_s/\delta t$ and $\delta E_s/\delta t$,
     respectively.
Moreover, $\vec{q}_s$ is the heat flux carried by species $s$, and
     $Q_s$ stands for the heating rate applied to
     species $s$ from non-thermal processes.
In equation~(\ref{gen_vec_vk}), the subscript $j$ stands for ion species other than $k$,
     namely, $j=p$ for $k=\alpha$ and vice versa.
As can be seen, in addition to
     the term $(\nabla\times \vec{B})\times\vec{B}$, the Lorentz force possesses 
     a new term in the form of the cross product of the ion velocity difference and
     magnetic field.
Physically, this new term represents the mutual gyration of one ion species about the other,
     the axis of gyration being in the direction of the instantaneous magnetic field.

Equations~(\ref{gen_nk}) to (\ref{gen_vec_magind}) form a complete set if 
     supplemented with the description of species heat fluxes $\vec{q}_s$
     and heating rates $Q_s$.
As such, they can be invoked to depict self-consistently the interaction
     between Alfv\'en waves and the solar wind species
     by explicitly introducing these waves via boundary conditions.
On the one hand solving this set of equations presents a computationally
     formidable task;
     on the other hand, one can extract the necessary
     information concerning the dynamical feedback of
     the waves to the plasma by going beyond the WKB limit. 
In the non-WKB approach to be adopted here, one assumes that the governing equations can still be
     separated into those governing the background time-independent flow and those 
     governing the transport of waves.
As noted by \citet{Lou_93} (also see the discussion), this separation does not necessarily require
     the waves be linear as long as 
     sufficiently small wave amplitudes are imposed at the Sun.

Further simplification also results from the choice of a flux tube coordinate
     system, in which the base vectors are 
     $\{\hat{e}_l, \hat{e}_N, \hat{e}_\phi\}$, 
     where
\begin{eqnarray*}
\hat{e}_l = \vec{B}_P/B_P, \hspace{0.5cm} \hat{e}_N = \hat{e}_\phi\times\hat{e}_l, 
\label{coor}
\end{eqnarray*}
     with the subscript $P$ denoting the poloidal component.
Moreover, the independent variable $l$ is the arclength along the
     poloidal magnetic field line.
This choice permits the decomposition of the magnetic field
     and species velocities into background ones and
     fluctuations, 
\begin{eqnarray}
\vec{B} = B_l \hat{e}_l + b \hat{e}_\phi,\hspace{0.5cm}
\vec{v}_s = U_s\hat{e}_l + w_s\hat{e}_N + u_s\hat{e}_\phi .
\label{vec_BV_comps}
\end{eqnarray}
     where $s=e, p, \alpha$. 
From the assumption of azimuthal symmetry, and the assumption that $\vec{B}_P$
     is time-independent,
     one can see from the poloidal component of equation~(\ref{gen_vec_magind})
     that $\vec{v}_{eP}$ should be strictly in the direction of $\vec{B}_P$.
In other words, $w_{e}=0$ to a good approximation.
Now let us consider the $\phi$ component of the momentum equation~(\ref{gen_vec_vk}).
Since the wave frequencies in question as well as other frequencies
     associated with the spatial dependence
     are well below the ion gyro-frequency
     $\Omega_k = (Z_k e B_l)/(m_k c)$ ($k=p, \alpha$), 
     from an order-of-magnitude estimate one can see that
     $|w_{j}-w_{k}| \ll |u_{k}|$.
Combined with the fact that $w_{e}=0$, this leads to that
     both $w_{p}$ and $w_{\alpha}$ should be very small and 
     can be safely neglected unless they appear alongside the
     ion gyro-frequency.
With this in mind, one can find from the $N$ component of equation~(\ref{gen_vec_vk}) that
\begin{eqnarray}
u_{\alpha}-u_{p}=\frac{b}{B_l}\left(U_{\alpha}-U_{p}\right). 
\label{ion_vdiff}
\end{eqnarray}
That is, the ion velocity difference is aligned with the instantaneous
     magnetic field.
This alignment condition further couples one ion species to the other.
Note that due to the assumption of quasi-zero current, equation~(\ref{ion_vdiff}) leads to
\begin{eqnarray}
u_k = u_e + \frac{b}{B_l} \left(U_k - U_e\right)
\label{ukandue}
\end{eqnarray}
    where $k=p, \alpha$. 

Given the aforementioned assumptions, the time-independent multicomponent solar wind
     in which the toroidal Alfv\'en waves propagate is governed by
\begin{eqnarray}
&& B_l\left(\frac{n_k U_k}{B_l}\right)' =0, 
      \label{bkg_nk}\\
&& U_{k}U_{k}' +\frac{p_k'}{n_k m_k} 
       +\frac{Z_k p_e'}{n_e m_k}+\frac{G M_\odot}{r} (\ln r)'   \nonumber \\
&&     -\frac{1}{n_k m_k}\left(\frac{\delta M_{k l}}{\delta t}
             +\frac{Z_k n_k}{n_e}\frac{\delta M_{e l}}{\delta t}\right)
       = a_{w,k}, \label{bkg_vk} \\
&& U_s \left(\frac{p_s}{\gamma -1}\right)'
       + \frac{\gamma p_s }{(\gamma-1)} B_l\left(\frac{U_{s}}{B_l}\right)'  \nonumber \\
&&     + B_l\left(\frac{q_s}{B_l}\right)' 
          -\frac{\delta E_s}{\delta t} -Q_s =0,  
               \label{bkg_ps} 
\end{eqnarray}
     where the prime denotes the derivative with respect to the arclength $l$
     which becomes the only independent spatial variable.
In addition, $a_{w,k}$ denotes the acceleration exerted on ion species $k$
     ($k=p, \alpha$) by the toroidal fluctuations,
\begin{eqnarray}
a_{w, k} &=& \left<u_{k}^2\right> (\ln R)'   
    -\frac{Z_k}{4\pi n_e m_k}\left <b\frac{\partial b}{\partial l}+b^2 (\ln R)'\right> \nonumber\\
    &-& \frac{\left<b X_k\right>}{B_l},
\label{def_wave_acce}
\end{eqnarray}
    where $R=r\sin\theta$ is the distance from
    a point along the poloidal magnetic field line to the magnetic axis.
Besides, the angular brackets stand for the time-average over one wave period.
The variable $X_k=\Omega_k (w_j-w_k)(Z_j n_j/n_e)$ distinguishes the present study
    from those using the ideal MHD  in which case
    $X_k \equiv 0$ \citep[e.g.,][]{HO80}. 
Apart from this, the wave-induced acceleration $a_{w,k}$ includes 
    the inertial centrifugal acceleration (the first term on the right hand side),
    and the usual $(\nabla\times \vec{B})\times\vec{B}$ term 
    (the second term).

\subsection{Transport of Toroidal Alfv\'en Waves}
\label{sec_waveeq}

The transport of toroidal Alfv\'en waves is governed by the azimuthal component
     of the momentum equation~(\ref{gen_vec_vk}) for ion species $k$ ($k=p, \alpha$)
     together with that of the magnetic induction law~(\ref{gen_vec_magind}).
To be more specific, these equations read
\begin{eqnarray}
    \frac{\partial b}{\partial t}
&+& B_l R \frac{\partial}{\partial l}
         \left[\frac{1}{R}\left(\frac{U_e}{B_l}b - u_e\right)\right] 
    = 0, \label{bphi} \\
    \frac{\partial u_k}{\partial t}
&+& U_k\left[\frac{\partial u_{k}}{\partial l}  + u_{k} (\ln R)'\right] \nonumber\\
&-&\frac{Z_k}{4\pi n_e m_k}B_l\left[
         \frac{\partial b}{\partial l} +b (\ln R)'\right] 
    = X_k. \label{vkphi1}  
\end{eqnarray}

The wave propagation is characterized by several key parameters, namely,
     the wave energy and energy flux densities
     as well as the wave-induced acceleration.
These parameters can be found by considering the energy conservation for the
     Alfv\'en waves,
\begin{eqnarray}
&&   \frac{\partial}{\partial t} 
         (\sum_k \frac{\rho_k u_k^2}{2} + \frac{b^2}{8\pi})  %%\nonumber \\
     +B_l\frac{\partial}{\partial l}
       \left\{\frac{1}{B_l}
             \left[\sum_k\frac{\rho_k U_k u_k^2}{2} \right. \right. \nonumber \\
&+&    \left.\left. \frac{b}{4\pi}\left(U_{e}b- B_l u_e\right) 
             \right]
       \right\} \nonumber \\
&=& -\sum_k \rho_k U_k u_k^2(\ln R)' %\nonumber \\
    +\frac{U_e}{4\pi} 
         \left[b\frac{\partial}{\partial l}b
            + b^2(\ln R)' \right] \nonumber \\
&&     +\sum_k \rho_k u_k X_k .
\label{enercons}
\end{eqnarray}
On the left hand side (LHS), the first term is the time derivative of the
     perturbation energy density,
     while the second is the divergence of the perturbation flux density.
The physical meaning of the right hand side (RHS)
     can be revealed as follows.
By using relation~(\ref{ukandue}) one finds 
\begin{eqnarray*}
   \sum_k \rho_k u_k X_k -\sum_k \rho_k U_k \frac{b X_k}{B_l} 
 = \left(u_e-\frac{b}{B_l}U_e\right) \sum_k \rho_k X_k .  
\end{eqnarray*}
Since $\sum_k \rho_k X_k = 0$, one can identify the time-average of the
     RHS of equation~(\ref{enercons}) as the negative of the work done by
     the wave-induced forces on
     the solar wind, i.e., $\sum_k \rho_k U_k a_{w, k}$ (cf. Eq.(\ref{def_wave_acce})).
In other words, taking the time-average of equation~(\ref{enercons}) yields
\begin{eqnarray}
B_l\frac{\partial}{\partial l} \frac{F_w}{B_l} = -\sum_k \rho_k U_k a_{w, k}, 
\label{enercons_aver}
\end{eqnarray}
     where $F_w$ is the time-average of the perturbation flux density and will
     be termed wave flux density for simplicity.
(Actually, the wave properties to be discussed always refer to time-averaged values.)
The total energy is therefore conserved for the system
     comprised of a multi-fluid solar wind and toroidal Alfv\'en waves:
     The gain in the solar wind kinetic energies is at the expense of the wave energy.

The appearance of $X_k$ makes equation~(\ref{vkphi1}) inconvenient to work with.
Instead one may consider the azimuthal component of
     the total momentum to eliminate $X_k$,
     the resultant equation being
\begin{eqnarray}
&& \sum_k \rho_k 
       \left[\frac{\partial u_k}{\partial t} 
            +\frac{U_k}{R}\frac{\partial}{\partial l}
            \left(R u_k\right)
       \right]  \nonumber \\
  &-&\frac{B_l}{4\pi R}\frac{\partial}{\partial l}\left(Rb\right) = 0.
\label{totvphi}
\end{eqnarray}
Now one may proceed by introducing the
     Fourier amplitudes at a given angular frequency $\omega$,
\begin{eqnarray}
[b(l, t), u_s (l, t)] = [\tilde{b}(l), \tilde{u}_s(l)] \exp(-i\omega t), \label{def_fourier}
\end{eqnarray}
    in which $s=e, p, \alpha$.
As a result, the Faraday's law~(\ref{bphi}) and equation~(\ref{totvphi}) now take the form
\begin{eqnarray}
&&  U_e \tilde{b}'-B_l \tilde{u}_e'   \nonumber \\
&=& \left[i\omega +U_e\left(\ln\frac{R B_l}{U_e}\right)'\right]\tilde{b} 
   -B_l(\ln R)' \tilde{u}_e , \label{induc_w} \\  
&& \left(\sum_k \rho_k U_k \frac{U_k-U_e}{B_l}-\frac{B_l}{4\pi}\right)\tilde{b}'
   + \sum_k \rho_k U_k \tilde{u}_e'  \nonumber \\
&=& -\left[\sum_k \frac{\rho_k U_k}{R}\left(R\frac{U_k-U_e}{B_l}\right)' 
          -\frac{B_l}{4\pi}\left(\ln R\right)'\right]\tilde{b} \nonumber\\
&-& \sum_k \rho_k U_k\left(\ln R\right)'\tilde{u}_e \nonumber\\
&+& i\omega\sum_k\rho_k\left(\frac{U_k-U_e}{B_l}\tilde{b}+\tilde{u}_e\right) ,
\label{totmomen_w}
\end{eqnarray}
    in which we have used relation~(\ref{ukandue})
    to express the ion velocity fluctuation
    $\tilde{u}_k$ in terms of the electron one $\tilde{u}_e$.

Once the background flow parameters and proper boundary conditions are given, 
    equations~(\ref{induc_w}) and (\ref{totmomen_w})
    can be solved for the Fourier amplitudes of magnetic fluctuation $\tilde{b}$ and
    the electron velocity fluctuation $\tilde{u}_e$ for a given angular frequency $\omega$.
The ion velocity fluctuations $\tilde{u}_k$ can
    then be found in virtue of relation~(\ref{ukandue}).
The three wave-related parameters can be evaluated by forming a time-average
    over one wave period.
Specifically, the wave energy density $E_w$ and wave energy flux density $F_w$ are
    given by
\begin{subequations}
\label{def_EFa_wave}
\begin{eqnarray}
E_w &=& \sum_k \frac{\rho_k \left<u_k^2\right> }{2} 
        +\frac{\left<b^2\right>}{8\pi}, \label{def_Ew} \\
F_w &=& \sum_k \frac{\rho_k U_k \left<u_k^2\right>}{2}   \nonumber\\
   &+&\frac{1}{4\pi}\left(U_e \left<b^2\right>-B_l \left<b u_e\right>\right), \label{def_Fw}
\end{eqnarray}
\end{subequations}
     while the wave-induced acceleration $a_{w,k}$ has already been given by
     equation~(\ref{def_wave_acce}).
It can be seen that $E_w$ consists of the ion kinetic 
     as well as the magnetic energies,
     while $F_w$ is comprised of the kinetic energy flux 
     convected by ion fluids and the Poynting flux.
To evaluate the time-average $\left<f g\right>$ for two wave-related fluctuations
     $[f,g] \sim [\tilde{f},\tilde{g}] \exp(-i\omega t)$,
     one may use the expression \citep[cf.][]{MacChar_94}
\begin{eqnarray}
\left<f g\right> = \frac{1}{2}\mathrm{Re}(\tilde{f} \tilde{g}^*)
      = \frac{1}{4}(\tilde{f} \tilde{g}^* + \tilde{f}^* \tilde{g}),
\label{def_average}
\end{eqnarray}
     where $\mathrm{Re}(z)$ denotes the real part of a complex variable $z$, and 
     the superscript asterisk denotes the complex conjugate.

Although in general one has to rely on a numerical integrator to solve
    equations~(\ref{induc_w}) and (\ref{totmomen_w}),
    as shown below, they are analytically tractable for two limiting cases.

\section{ANALYTICAL SOLUTIONS IN TWO LIMITS}
\label{sec_ana_sol}

In this section, we will examine the analytical solutions to
     equations~(\ref{induc_w}) and (\ref{totmomen_w}) 
     in the WKB and zero-frequency ($\omega = 0$) limits.
These analytical treatments not only help to validate numerical solutions,
     but also allow one to gain insights into the mathematical properties of
     the governing equations.

\subsection{the WKB Limit}
\label{sec_wkb}

Extensive studies have been made on the Alfv\'en waves in the WKB limit.
In particular, the Alfv\'en wave force exerted on differentially streaming ions
     was first derived by \citet{Hollweg_74}
     and later by \citet{McKenzie_etal_79} for a spherically expanding solar wind.
Based on the idea of the wave average Lagrangian, the work by \citet{IH82} 
     not only obtained the expression for the wave force but further
     showed that there exists an adiabatic invariant, namely the wave action flux,
     in the absence of wave dissipation.
The derivation is rather general and does not require a specific flux tube geometry.
Alternatively, the action flux conservation has been independently obtained
     by \citet{McKenzie_94} who used the more familiar WKB analysis.
Although \citet{McKenzie_94} assumed that the solar wind is again spherically symmetric,
     his results can be readily extended to an axisymmetric configuration.
In what follows, such an extension is presented.

The formal development of the WKB analysis starts with the introduction of
    the expansion
\begin{subequations}
\label{def_wkb_expan}
\begin{eqnarray}
&\tilde{u}_s = (u_{s, 1} + u_{s, 2} +\cdots) \exp(iS(l)), \\
&\tilde{b} = (b_1 + b_2 + \cdots) \exp(iS(l)),
\end{eqnarray}
\end{subequations}
     in which $s= e, p, \alpha$.
It is assumed that $u_{s, n}$ and $b_n$ ($n=1, 2, ...$) vary at
     the same spatial scale $H$ as the background flow parameters,
     and $1/H$ is small compared with the wave number $K=S'(l)$, i.e.,
     $\mu = 1/(K H) \ll 1$.
In addition, it is assumed that $|u_{s, n+1}/u_{s, n}|\sim |b_{n+1}/b_{n}|\sim \mu$.
Substituting the expansion~(\ref{def_wkb_expan}) into 
     equations~(\ref{induc_w}) and (\ref{totmomen_w}), one can find by 
     sorting different terms according to orders of $\mu$ that
\begin{subequations}
\label{wkb_eqs}
\begin{eqnarray}
&&\omega_e b_n + K B_l u_{e,n} \nonumber \\
&=& -i R B_l\left[\frac{1}{R}\left(\frac{U_e}{B_l}b_{n-1}
         -u_{e, n-1}\right)\right]' , \label{wkb_ind} \\
&& \sum_k \rho_k \omega_k \left(u_{e, n} + \frac{U_k-U_e}{B_l}b_n\right)
       + K\frac{B_l}{4\pi} b_n  \nonumber\\
&=& -i\sum_k \frac{\rho_k U_k}{R}\left[
         R\left(u_{e, n-1}+\frac{U_k-U_e}{B_l} b_{n-1}\right)\right]' \nonumber \\
&+&i \frac{B_l}{4\pi R}\left(R b_{n-1}\right)', \label{wkb_mom}
\end{eqnarray}
\end{subequations}
     where the definition $\omega_s = \omega - K U_s$ ($s=e, p, \alpha$) has been used.

At the lowest order $n=1$, the RHS of equation~(\ref{wkb_eqs})
     is zero.
For $u_{e, 1}$ and $b_1$ not to be identically zero, the determinant of 
     the coefficient matrix has to be zero.
As a result, one finds the local dispersion relation
\begin{eqnarray}
\sum_k \rho_k (U_{ph}-U_k)^2 = \frac{B_l^2}{4\pi}, 
\label{wkb_disprel}
\end{eqnarray}
     where $U_{ph}=\omega/K$ is the local phase speed of Alfv\'en waves. 
Furthermore, one can find the local eigen-relations at order~$n=1$, 
\begin{eqnarray}
u_{s, 1} = -\left(U_{ph}-U_s\right)\frac{b_1}{B_l} ,
\label{wkb_eigen1}
\end{eqnarray}
     with $s= e, p, \alpha$.

At an arbitrary order $n\ge 2$, it follows from equation~(\ref{wkb_ind})  that
\begin{eqnarray*}
u_{e, n} = - \left(U_{ph}-U_e\right)\frac{b_{n}}{B_l}  + \bar{u}_n,
\label{wkb_eigenn}
\end{eqnarray*}
    in which
\begin{eqnarray*}
\bar{u}_n =-\frac{i R}{K}\left[\frac{1}{R}\left(\frac{U_e}{B_l}b_{n-1}-u_{e, n-1}\right)\right]'
\end{eqnarray*}
    arises from quantities at order $n-1$.
At order $n+1$, one can eliminate $b_{n+1}$
    and $u_{e, n+1}$ on the LHS of equation~(\ref{wkb_eqs}) by
    using the dispersion relation~(\ref{wkb_disprel}),
    the resultant equations taking the form
\begin{eqnarray}
T_n + S_n &=& 0, 
\end{eqnarray}
    where
\begin{subequations}
\label{def_SnTn}
\begin{eqnarray}
S_n &=& -\left[\sum_k \rho_k \omega_k R B_l \left(\frac{\bar{u}_n}{R}\right)'\right. \nonumber \\
    &+& \left. K B_l \sum_k \frac{\rho_k U_k}{R}\left(R\bar{u}_n\right)'\right] , 
            \label{def_Sn}   \\
T_n &=& \sum_k \rho_k \omega_k R B_l \left(\frac{U_{ph} b_n}{R B_l}\right)' \nonumber \\
    &+&K B_l \sum_k \frac{\rho_k U_k}{R} \left(R\frac{U_{ph}-U_k}{B_l} b_n\right)' \nonumber \\ 
    &+&K \frac{B_l^2}{4\pi R}\left(R b_n\right)'. 
            \label{def_Tn}
\end{eqnarray}
\end{subequations}
The barred and unbarred parts are again distinguished.

On expanding $T_n$, by repeatedly using the dispersion relation~(\ref{wkb_disprel}) one finds that
\begin{eqnarray}
 T_n=\frac{K B_l^3}{U_{ph} b_n}\left[\left(\frac{U_{ph} b_n}{B_l}\right)^2
     \frac{\rho (U_{ph}-U_m)}{B_l}\right]',
 \label{wkb_Tn}
\end{eqnarray}
     in which the following definitions have been used \citep[cf.][]{McKenzie_94}
\begin{subequations}
\label{def_rhovmvj2}
\begin{eqnarray}
\rho = \sum_k \rho_k, \\
\rho U_m = \sum_k \rho_k U_k, \\
\rho U_j^2 = \sum \rho_k U_k^2, 
\label{def_rhovmvj}
\end{eqnarray}
\end{subequations}
     with the last one defined for future use.
If putting $n=1$, from $S_1=0$ one finds $T_1=0$, i.e., 
\begin{eqnarray}
b_1^2 \frac{\rho (U_{ph}-U_m) U_{ph}^2}{B_l^3} = \mbox{const}.
\label{wkb_action_txt}
\end{eqnarray}
Note that equation~(\ref{wkb_action_txt}) can be interpreted in terms of 
     the conservation of wave action flux, which was first found 
     by \citet{IH82}.
On the other hand, expanding equation~(\ref{vkphi1}) to first order in $\mu$
     and using the resultant $X_k$ to evaluate $\left< b X_k\right>$ 
     in equation~(\ref{def_wave_acce}), with the aid of
     the eigen-relation~(\ref{wkb_eigen1}), one can find 
     a compact expression for the wave-induced acceleration on 
     ion species $k$ ($k=p, \alpha$),
\begin{eqnarray}
a_{w, k} = \left(\frac{U_{ph}^2-U_k^2}{2 B_l^2} \left<b^2\right>\right)',
\label{wkb_acce}
\end{eqnarray}
     where the time-average $\left<b^2\right>=|b_1|^2/2$.
It is noteworthy that the shape of the line of force, which enters into
     the discussion through $R=r\sin\theta$,
     does not show up via $S_{n-1}$ unless one examines the evolution of higher order fluctuations
     $u_{s, n}$ and $b_n$ ($n\ge 2$).

{Some remarks on the wave acceleration $a_{w,k}$ given by Equation~(\ref{wkb_acce}) are necessary.
One may find that
    $(U_\alpha^2-U_p^2)(1+|b_1|^2/(2B_l^2))=\mbox{const}$
    if the flow speeds are entirely determined by the wave forces.
Usually $|b_1|/B_l$ increases with increasing distance. 
The net effect of $a_{w,k}$ is therefore to limit the proton-alpha differential speed
    $U_{\alpha p}$.
In this sense $a_{w,k}$ was likened to an additional frictional force
     \citep[e.g.,][]{Hollweg_74}.
As noted by \citet{Hollweg_74}, the differential nature of $a_{w,k}$ derives eventually from
     a wave-induced transverse drift velocity $\delta\vec{u}_{k}$ ($k=p, \alpha$),
     which is different for different species because they see different wave electric fields
     when they flow differentially.
The contribution to the wave force from $\delta\vec{u}_{k}$ is twofold.
First, it induces a centrifugal force that is always non-negative.
Second, it contributes to the species electric current flow $\delta\vec{J}_{\perp,k}$
     which in turn exerts on ion species $k$ a $\delta\vec{J}_{\perp,k} \times \delta\vec{B}$ force,
     where $\delta\vec{B}$ is the wave magnetic field.
This force may become negative.
As a matter of fact, the two terms
     may even add up to a negative value: The wave force
     may decelerate rather than accelerate the ion species.
Consider a simplified situation similar
     to that considered by \citet{Hollweg_74}.
One assumes that
     the spatial variation of both $U_\alpha$
     and $U_p$ can be neglected,
     and $U_\alpha > U_p > U_A$ where $U_A=B_l/\sqrt{4\pi\rho}$
     is the bulk Alfv\'en speed.
One may further assume that the alpha particles are test particles, 
     the background magnetic field is purely radial,
     and that $|b_1|^2 \propto r^{-\epsilon}$ with $\epsilon$ being a positive constant.
In this case, from Equation~(\ref{wkb_acce}) one finds that
\begin{eqnarray*}
a_{w,\alpha} =\frac{|b_1|^2}{4B_l^2 r}\left\{
\left[\left(4-\epsilon\right)U_p^2+\epsilon U_A^2\right]
   -\left(4-\epsilon\right)\left(U_{\alpha p}+U_p\right)^2
\right\} .
\end{eqnarray*}
It follows that $a_{w,\alpha} < 0$ when $U_{\alpha p} > U_c$ where
      the critical value $U_c$ can be roughly approximated by
\begin{eqnarray*}
 U_c \approx \frac{\epsilon}{2(4-\epsilon)}\frac{U_A}{U_p} U_A,
\end{eqnarray*}
      which may be smaller than $U_A$
      by a factor of ${U_A}/{U_p}$.
Noting that now the wave phase speed is $U_{ph}=U_p+U_A$, one finds that
      the waves may decelerate alpha particles even though $U_\alpha < U_{ph}$.
It turns out that this much simplified picture roughly represents the 
      behavior of $a_{w,\alpha}$ in the region between 20 and 30~$R_\odot$
      where $\epsilon$ can be taken to be $2.5$
      for the adopted background flow parameters
      (see Figure~\ref{fig_aw}b which indicates that $a_{w,\alpha}$
       in the WKB case becomes negative beyond 22.7~$R_\odot$ where $U_{\alpha p}$ exceeds $U_c$). 
}

\subsection{the Zero-frequency Limit}
\label{sec_zero_freq}

The other extreme will be $\omega = 0$. 
In this case, equations~(\ref{bphi}) and (\ref{totvphi}) 
     can be integrated to yield
\begin{subequations}
\label{zero_intg}
\begin{eqnarray}
&& u_e -U_e\frac{b}{B_l} = A_\Omega R,  \label{zero_corot}\\
&& R \left[\sum_k\frac{\rho_k U_k}{\rho U_m} u_k
     -\frac{B_l b}{4\pi \rho U_m}\right] 
   = A_L, \label{zero_totang}
\end{eqnarray}
\end{subequations}
     where $A_\Omega$ and $A_L$ are two integration constants.
When deriving equation~(\ref{zero_totang}), we have used the fact that $\rho U_m/B_l$ is a constant.
Equation~(\ref{zero_intg}) demonstrates a clear connection to the problem
     of angular momentum transport in a multicomponent solar wind, as has been
     discussed by \citet{LiLi06}.

With the aid of equation~(\ref{ukandue}), equation~(\ref{zero_intg}) can
     be solved to yield
\begin{eqnarray}
b = R \frac{4\pi\rho U_m}{B_l}\frac{A_L/R^2-A_\Omega}{M_T^2-1},
\label{zero_b_prem}
\end{eqnarray}
     where 
\begin{eqnarray}
M_T^2 = \frac{4\pi\rho U_j^2}{B_l^2}
\label{def_mt2}
\end{eqnarray}
     is the square of the combined Alfv\'en Mach number.
For typical solar winds, there exists one point where $M_T=1$ between 1~$R_\odot$
     and 1~AU.
Call this point the Alfv\'enic point, and let it be 
     denoted by subscript $A$.
It then follows that for $b$ not to be singular at this point, one must require
     $A_L = A_\Omega R_A^2$.
As a result, one may obtain
\begin{subequations}
\label{zero_exprs}
\begin{eqnarray}
b = && A_\Omega R \frac{4\pi\rho U_m}{B_l}\frac{(R_A/R)^2-1}{M_T^2-1} , 
          \label{zero_b} \\
u_s =&& A_\Omega R\left[1+\frac{4\pi\rho U_m U_s}{B_l^2}
       \frac{(R_A/R)^2-1}{M_T^2-1}\right]
          \label{zero_us}, 
\end{eqnarray}
\end{subequations}
     where $s=e,p,\alpha$.
 
\section{NUMERICAL MODEL AND METHOD OF SOLUTION}
\label{sec_num_model}

Apart from the WKB and zero-frequency limits, equations~(\ref{induc_w})
     and (\ref{totmomen_w}) can be integrated only numerically.
Nevertheless, the treatment in the zero-frequency limit in the previous section
     reveals that the Alfv\'enic point where $M_T^2=1$ is a singular point
     for the system of equations.
One may expect that this is also the case for an arbitrary finite $\omega$.
In their present form, however, equations~(\ref{induc_w})
     and (\ref{totmomen_w}) do not show explicitly the existence of such 
     a singular point.
Hence they are not convenient to work with numerically and
     need to be further developed in what follows.
Moreover, we will also describe how to specify the background flow parameters, 
     and the solution procedure as well.

\subsection{Further Development of Governing Equations}
\label{sec_gen_num}

For the convenience of presentation, one may define, in addition to $M_T$,
     the following dimensionless parameters 
\begin{eqnarray}
&& X_{me} = \frac{4\pi \rho U_m U_e}{B_l^2}, 
X_{ee} = \frac{4\pi \rho U_e U_e}{B_l^2}, \nonumber\\
&& X_{eA} = \frac{4\pi \rho U_e U_A}{B_l^2}, 
X_{mA} = \frac{4\pi \rho U_m U_A}{B_l^2}, \nonumber\\
&& Z = M_T^2-1-X_{me}, 
\label{def_xandz}
\end{eqnarray}
     in which $U_A = B_l/\sqrt{4\pi\rho}$
     is the bulk Alfv\'en speed.
Now equation~(\ref{totmomen_w}) can be expressed as
\begin{eqnarray}
Z\tilde{b}'
   &+& \frac{4\pi \rho U_m}{B_l}\tilde{u}_e'  
= -\frac{(R Z)'}{R} \tilde{b}
    -\frac{4\pi\rho U_m (\ln R)'}{B_l} \tilde{u}_e \nonumber \\
&& +i\omega\left(\frac{4\pi\rho (U_m -U_e)}{B_l^2}\tilde{b}
   +\frac{4\pi\rho}{B_l}\tilde{u}_e\right) . \label{num_bu1} 
\end{eqnarray}
Taking into account equation~(\ref{induc_w}), one finally arrives at
 \begin{subequations}
\label{num_eqs}
 \begin{eqnarray}
 \left(M_T^2-1\right){\xi}' &=& F_{11} \xi +F_{12} \eta ,\\
\left(M_T^2-1\right){\eta}' &=& F_{21} \xi +F_{22} \eta ,
\end{eqnarray}
\end{subequations}
      where two dimensionless variables have been introduced for convenience,
\begin{eqnarray}
\xi = \tilde{b}/B_l, \hspace{0.4cm} \eta=\tilde{u}_e/U_A.
\label{def_xi_eta}
\end{eqnarray}
Moreover, the coefficients are
\begin{subequations}
\label{def_num_coefEs}
\begin{eqnarray}
F_{11} 
      &=& \frac{i \omega}{U_e}\left(2X_{me}-X_{ee}\right) 
          + X_{me}\left(\ln \frac{R}{U_e} \right)' \nonumber \\
      &&  -\frac{(RB_l Z)'}{RB_l} \\
F_{12} 
      &=& \frac{i \omega}{U_e}X_{eA}
              -2X_{mA}\left(\ln R\right)'  \\
F_{21}
      &=& -\frac{i \omega}{U_A}\left(Z-X_{me}+X_{ee}\right) \nonumber \\
      &-&  \frac{U_e}{U_A}\left[Z\left(\ln \frac{R B_l}{U_e}\right)' + \frac{(RZ)'}{R}
            \right] \\
F_{22} &=& \frac{i \omega}{U_e}X_{ee}+\left(Z-X_{me}\right)\left(\ln R\right)' \nonumber\\
       &-& (M_T^2-1)(\ln U_A)'.
\end{eqnarray}
\end{subequations}

It is now clear that the Alfv\'enic point where $M_T^2=1$ is a singular point for
    equation~(\ref{num_eqs}).
The ensuing task is therefore to find solutions that pass smoothly through
    the Alfv\'enic point.

\subsection{Background Three-fluid Solar Wind}
\label{sec_background}

In principle, one needs to solve equations~(\ref{bkg_nk}) to (\ref{bkg_ps}) to find 
     a realistic background solar wind by neglecting the wave-induced acceleration
     in equation~(\ref{bkg_vk}) and using a suitable heat input $Q_s$ ($s=e, p, \alpha$).
However, it is observationally established that in the heliocentric 
     range $r> 0.3$~AU, the proton-alpha differential speed
     $U_{\alpha p} = U_\alpha-U_p$ 
     closely tracks the local Alfv\'en speed in the fast solar wind with 
     $U_p > 600$\velunits \citep{Marsch_etal_82}.
So far this fact still poses a theoretical challenge: Adjusting the heating parameters
     proves difficult to yield such a behavior for $U_{\alpha p}$.
In what follows, we shall adopt a two-step approach to find the background flow parameters.
First, equations~(\ref{bkg_nk}) to (\ref{bkg_ps}) are solved by using a suitable set of
     heating parameters along a prescribed meridional magnetic field line.
As a result, we have $\rho_p$, $U_p$, $(\rho_\alpha U_\alpha)_I$ and 
     $U_{\alpha p, I}$ for the whole heliocentric range
     from 1~$R_\odot$ out to 1~AU.
Note that the subscript $I$ denotes the results obtained from this step.
Second, the alpha parameters are altered as follows.
We first impose an {\it ad hoc} profile for $U_{\alpha p}$ which is identical to
     $U_{\alpha p, I}$ in the range $r\lesssim 0.3$~AU
     but undergoes a smooth transition to a profile that
     is roughly 0.8~$U_A$ everywhere for $r\gtrsim 0.3$~AU.
The alpha speed and mass density are then given by
    $U_\alpha = U_p + U_{\alpha p}$ and 
    $\rho_\alpha = (\rho_\alpha U_\alpha)_I/U_\alpha$, respectively.

For the meridional magnetic field, we will adopt an analytical model given by 
     \citet{Bana_etal_98}.
In the present implementation, the model magnetic field consists of
     dipole and current-sheet components only.
A set of parameters $M=3.6222$, $Q=0$, $K=1.0534$ and $a_1=2.5$ are
     chosen such that the last open magnetic field line is anchored
     at heliocentric colatitude $\theta=40^\circ$ on the Sun.

\begin{figure}
\begin{center}
\epsscale{1.15}
\plotone{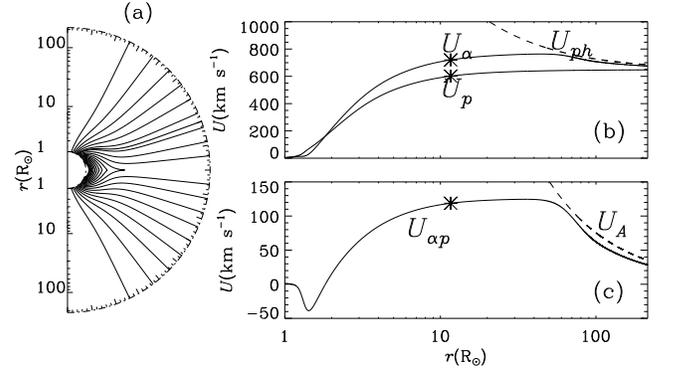}
\caption{Background 3-fluid solar wind in which the toroidal Alfv\'en waves propagate.
(a) Adopted poloidal magnetic field configuration extending from the 
     coronal base {(1~$R_\odot$)} to 1~AU{=215~$R_\odot$}. 
Note that the magnetic axis points upwards.
The thick contour delineates the line of force along which the wave equation is integrated.
(b) Proton and alpha flow speeds $U_p$ and $U_\alpha$ (solid lines),
      together with the 
      phase speed $U_{ph}$ (dashed line) expected in the WKB limit
      as given by equation~(\ref{wkb_disprel}).
(c) Proton-alpha differential speed $U_{\alpha p}=U_\alpha - U_p$ (solid line)
      and the local bulk Alfv\'en speed $U_A = B_l/\sqrt{4\pi\rho}$ (dashed),
      where $B_l$ is the poloidal magnetic field strength and $\rho$ the overall mass density.
The asterisks in panels (b) and (c) refer to the Alfv\'enic point where $M_T^2=1$, $M_T$
      being the combined Alfv\'en Mach number defined by equation~(\ref{def_mt2}).
 }
\label{fig_bkgrnd}
\end{center}
\end{figure}

The background magnetic field configuration and flow parameters
     are depicted in Figure~\ref{fig_bkgrnd}.
In Fig.\ref{fig_bkgrnd}a, the field line along which we
     integrate the solar wind
     equations~(\ref{bkg_nk}) to (\ref{bkg_ps}) and 
     the wave equation~(\ref{num_eqs})
     is delineated by the thick contour.
This line of force is rooted at colatitude
     $\theta=31.5^\circ$ on the Sun where the poloidal magnetic field strength
     $B_l$ is $6.6$~G.
It reaches $\theta=70^\circ$ at 1~AU where $B_l$ is 3.3$\gamma$, 
     compatible with the Ulysses measurements \citep{SmithBalogh_95}.
In Fig.\ref{fig_bkgrnd}b, the solid lines give the distribution
     with heliocentric distance $r$ of the proton ($U_p$) and alpha
     ($U_\alpha$) speeds,
     while the dashed curve is for the phase speed
     ($U_{ph}$) of Alfv\'en waves in the WKB limit 
     as determined by equation~(\ref{wkb_disprel}).
Furthermore, Fig.\ref{fig_bkgrnd}c shows the distribution with $r$ of 
     the proton-alpha differential
     speed $U_{\alpha p}$ (the solid line)
     and Alfv\'en speed $U_A$ (the dashed curve).
The asterisks in Figs.\ref{fig_bkgrnd}b and \ref{fig_bkgrnd}c
     denote the Alfv\'enic point where $M_T^2=1$, which is located
     at $r_A=11.6$~$R_\odot$ in the chosen background model.

It can be seen in Fig.\ref{fig_bkgrnd}b that $U_{ph}$ is much larger than the ion flow speeds
     below $\sim 20$~$R_\odot$ but is close to $U_\alpha$ beyond
     $\sim 0.3$~AU.
This latter behavior is understandable since one can find from 
     equation~(\ref{wkb_disprel}) that
\begin{eqnarray}
   U_{ph} 
&=& U_m + \sqrt{U_A^2 - \frac{\rho_\alpha \rho_p}{\rho^2}U_{\alpha p}^2} \nonumber \\
&\approx& U_\alpha + U_A \left(1-x_p -\frac{x_p x_\alpha}{2}\right) ,
\label{wkb_vph}
\end{eqnarray}
     where $x_k = (\rho_k/\rho)(U_{\alpha p}/U_A)$ ($k=p, \alpha$).
The second expression is very accurate for $U_{\alpha p} \le U_A$ which is the case
     for the assumed background flow (cf. Fig.\ref{fig_bkgrnd}c).
Hence $U_{ph} \approx U_\alpha$ in the region $r\gtrsim 0.3$~AU where
     $U_\alpha > U_A$ and $x_p+x_p x_\alpha/2 \sim 1$.
Moreover, it can be seen from Fig.\ref{fig_bkgrnd}c that $U_{\alpha p}$ drops from 
     nearly zero at $R_\odot$ to $-38.7$\velunits\ at 1.42~$R_\odot$,
     beyond which $U_{\alpha p}$ increases gradually.
In interplanetary space $r\gtrsim 0.3$~AU, $U_{\alpha p}$ can be seen
     to follow $U_A$ closely, as is required.

At the Alfv\'enic point $r_A$, the proton speed $U_p$ is 602\velunits, the alpha one
     $U_\alpha$ is 721\velunits.
Eventually $U_p$ reaches 648\velunits\ at 1~AU where 
     $U_\alpha$ is 676\velunits.
As for the density parameters $n_p$ and $n_\alpha$, the model yields
     $n_p = 3.57$ cm$^{-3}$ and $n_\alpha/n_p = 4.6\%$
     at 1~AU.
As a result, the proton (alpha) flux $n_p U_p$ ($n_\alpha U_\alpha$)
     is 2.32 (0.11)$\times 10^8$~cm$^{-2}$~s$^{-1}$
     when scaled to 1~AU.
All these values are consistent with typical measurements for the low-latitude 
     fast solar wind streams,
     e.g., those made by Ulysses \citep{McComas_etal_00}.

\subsection{Solution Procedure and Boundary Conditions}
\label{sec_sol_proc}

Given the background flow parameters, 
     the coefficients in equation~(\ref{def_num_coefEs})
     can be readily evaluated for a given angular frequency $\omega$.
Equation~(\ref{num_eqs}) is ready to solve once
     proper boundary conditions are supplemented.
As is well known, one boundary condition has to be imposed at the Alfv\'enic point to ensure
     the solution is regular.
To establish this, let us consider the region adjacent to the Alfv\'enic point $l_A$, such that
     any function $f(l)$ can be Taylor-expanded,
\begin{eqnarray}
 f (l) = f^{(0)} + f^{(1)} (l-l_A),  \label{def_alf_expan}
\end{eqnarray}
     in which $f^{(0)} = f(l_A)$.
When substituting this expansion into equation~(\ref{num_eqs}), one finds
\begin{subequations}
\label{num_alfvpnt}
\begin{eqnarray}
&& F_{11}^{(0)} \xi^{(0)} + F_{12}^{(0)} \eta^{(0)} = 0, \label{num_alf_zero1} \\
&& F_{21}^{(0)} \xi^{(0)} + F_{22}^{(0)} \eta^{(0)} = 0, \label{num_alf_zero2} \\
&& F_{11}^{(0)} \xi^{(1)} + F_{11}^{(1)} \xi^{(0)}
        +F_{12}^{(0)} \eta^{(1)} + F_{12}^{(1)} \eta^{(0)} \nonumber\\
        &&= (M_T^2)^{(1)} \xi^{(1)}, \\
&& F_{21}^{(0)} \xi^{(1)} + F_{21}^{(1)} \xi^{(0)}
        +F_{22}^{(0)}\eta^{(1)}+F_{22}^{(1)}\eta^{(0)}  \nonumber\\
        &&= (M_T^2)^{(1)} \eta^{(1)} .
\end{eqnarray}
\end{subequations}
It is easy to recognize that equations~(\ref{num_alf_zero1}) and (\ref{num_alf_zero2})
     are not independent from each other, since at the Alfv\'enic point $l_A$,
     $F_{11}^{(0)}/F_{21}^{(0)} = F_{12}^{(0)}/F_{22}^{(0)}= (U_A/U_e) (l_A)$.
Hence at $l_A$, the unknowns
     $\xi^{(1)}$, $\eta^{(0)}$ and $\eta^{(1)}$ can all be expressed in terms of $\xi^{(0)}$,
     which can be arbitrarily chosen.
Once these four quantities are known, the wave quantities
     $\xi$ and $\eta$ can be obtained by the expansion~(\ref{def_alf_expan})
     at the two grid points immediately astride the Alfv\'enic point.
Equation~(\ref{num_eqs}) is then integrated by using 
     a fourth order Runge-Kutta method both inwards
     to the coronal base $R_\odot$ and outwards to 1~AU
     \citep[see][]{CranmerBalle_05}.
We then rescale the obtained solution such that 
     $|\tilde{u}_e|$ is fixed at $R_\odot$ 
     for all $\omega$ to be considered.

\section{NUMERICAL SOLUTIONS}
\label{sec_num_res}
In this section, the solutions to equation~(\ref{num_eqs}) with a number of different
     angular frequencies $\omega$ are presented.
All these solutions have the same amplitude for the electron velocity fluctuation
     $|\tilde{u}_e| = 10\sqrt{2}$\velunits, or equivalently
     the time average~$\left<u_e^2\right>^{1/2}=10$\velunits.
Note that such a value is only about $1/3$ of the upper limit
     of the nonthermal velocity amplitude derived from
     line width measurements \citep{Banerjee_etal_98}.
With this choice at the coronal base, we can avoid 
     awkwardly large wave amplitudes in interplanetary space,
     a natural consequence of the assumption that
     no wave dissipation is present.
Nevertheless, the base amplitude can be seen
     as a free scaling parameter for equation~(\ref{num_eqs}).
The relative deviation of the non-WKB from WKB results
     is independent
     from this choice, even though the absolute magnitudes are affected.

\begin{figure}
\begin{center}
\epsscale{1.2}
\plotone{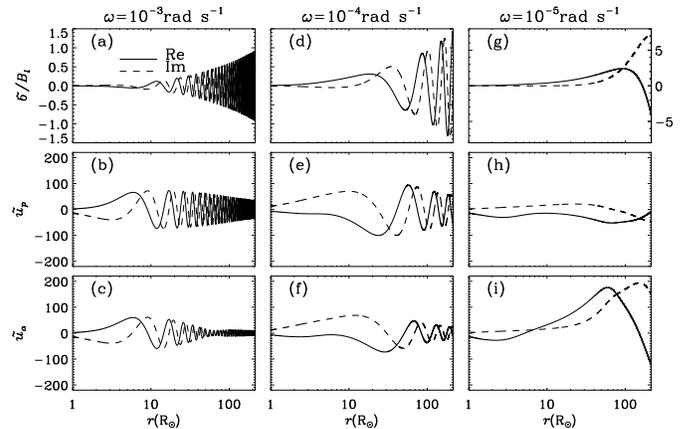}
\caption{
Radial profiles of the Fourier amplitudes.
Results are shown for the real (solid curves)
     and imaginary parts (dashed) for 
     three angular frequencies $\omega=10^{-3}$ (left column),
     $10^{-4}$ (middle) and $10^{-5}$\angfrequnits\ (right).
(a), (d) and (g) Magnetic fluctuation given in terms of $\tilde{b}/B_l$, $B_l$ being
     the background magnetic field strength.
(b), (e) and (h) Proton velocity fluctuation $\tilde{u}_p$.
(c), (f) and (i) Alpha velocity fluctuation $\tilde{u}_\alpha$.
Note that panel (g) uses a scale different from (a) and (d).
}
\label{fig_ReIm}
\end{center}
\end{figure}

Figure~\ref{fig_ReIm} presents the radial profiles of the real (solid lines) 
     and imaginary (dashed lines) parts of the Fourier amplitudes 
     with three different angular frequencies, $\omega=10^{-3}$ (left column), $10^{-4}$ (middle)
     and $10^{-5}$\angfrequnits (right).
Since the modulus of the magnetic fluctuation $|\tilde{b}|$
     spans several orders of magnitude,
     $\tilde{b}/B_l$ is plotted instead of $\tilde{b}$
     in panels (a), (d) and (g).
Note that panel (g) uses a scale different from (a) and (d).
Panels (b), (e) and (h) depict the proton velocity fluctuation $\tilde{u}_p$,
     while panels (c), (f) and (i) give the alpha one $\tilde{u}_\alpha$.

The most prominent feature of Figs.\ref{fig_ReIm}a, \ref{fig_ReIm}b and \ref{fig_ReIm}c
     for $\omega=10^{-3}$\angfrequnits\
     is that the radial dependence of the real and imaginary parts of fluctuations
     exhibits a clear oscillatory behavior.
This is particularly true beyond $\sim 10$~$R_\odot$.
Further inspection of the region $r \gtrsim 10$~$R_\odot$ indicates that the magnetic and
     fluid velocity fluctuations are well correlated, as would be expected
     from the eigen-relation~(\ref{wkb_eigen1})
     obtained in the WKB limit.
The WKB nature is further revealed by examining the phase relation,
     i.e., the displacement between the real and imaginary parts,
     for any of the three Fourier amplitudes.
For instance, for the $\tilde{u}_p$ profile the nodes in the real part
     correspond well to the troughs or crests in the imaginary part.
Now if examining the envelopes of the Fourier amplitude profiles, one can see that 
     $|\tilde{b}|/B_l$ (Fig.\ref{fig_ReIm}a) tends to increase with
     $r$ while $|\tilde{u}_p|$ tends to decrease beyond say 10~$R_\odot$.
On the other hand, $|\tilde{u}_\alpha|$ exhibits a non-monotonic behavior and possesses a
     local minimum at $\sim 68$~$R_\odot$. 
Moreover, the magnitude of $\tilde{u}_\alpha$
     is considerably smaller than that of
     $\tilde{u}_p$ beyond 0.3~AU.
This is understandable in light of equation~(\ref{wkb_eigen1}) since
     the local phase speed of Alfv\'en waves $U_{ph}$ is close
     to the alpha flow speed $U_\alpha$ (see Fig.\ref{fig_bkgrnd}b).

The Fourier amplitudes for $\omega=10^{-4}$\angfrequnits\ (the middle column)
     also possess an oscillatory behavior, however the {exact} correlation
     between the magnetic (Fig.\ref{fig_ReIm}d)
     and ion velocity (Figs.\ref{fig_ReIm}e and \ref{fig_ReIm}f) fluctuations is gone,
     as is the {exact} phase relation between the real and imaginary
     parts.
Furthermore, one can see that the magnitude of the alpha velocity
     fluctuations (Fig.\ref{fig_ReIm}f) in interplanetary space
     is substantially larger than that for $\omega=10^{-3}$\angfrequnits.
Now let us move on to the right column for which $\omega=10^{-5}$\angfrequnits.
Distinct from the preceding two columns, the oscillatory feature disappears 
     altogether. 
Instead, the real and imaginary parts of the Fourier amplitudes evolve slowly
     with radial distance $r$.
We note that such a transition from wave-like to quasi-steady dependence
     on $r$ with decreasing $\omega$ was explored in detail
     by \citet{HO80} \citep[see also][]{Lou_93, MacChar_94}.
That there exists a critical frequency below which the transition occurs 
     was interpreted in terms of the coupling between inwardly
     and outwardly propagating waves.
Although in the present study, the propagation in opposite directions
     has not been explicitly
     separated, one can see that for a realistic 3-fluid solar wind,
     there also exists a similar critical frequency which may
     be approximated by 
\begin{eqnarray}
\omega_c \approx U_{mA} / (2 r_A) 
\label{def_omegac} 
\end{eqnarray}
     \citep[cf. Eq.(55) of][]{HO80}.
For the chosen background flow, we find $\omega_c \approx 3.8\times 10^{-5}$\angfrequnits.
The numerical experiments have confirmed the validity of this approximation.

\begin{figure}
\begin{center}
\epsscale{0.8}
\plotone{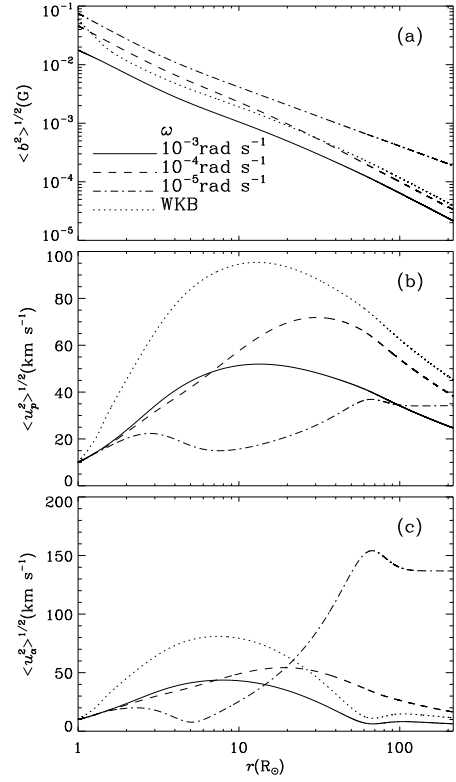}
\caption{
Time-average of the wave-related fluctuations. 
(a) Magnetic fluctuation $\left<b^2\right>^{1/2}$;
(b) and (c) Proton and alpha velocity fluctuations
           $\left<u_p^2\right>^{1/2}$ and $\left<u_\alpha^2\right>^{1/2}$.
Numerical results for three angular frequencies as well as for the WKB estimates
    are given in different line styles as indicated in panel (a).
}
\label{fig_ampltd}
\end{center}
\end{figure}

Figure~\ref{fig_ampltd} shows the radial evolution of the time-averages of 
     the magnetic $\left<b^2\right>^{1/2}$ as well as the fluid velocity fluctuations 
     $\left<u_k^2\right>^{1/2}$ ($k=p, \alpha$) for
     the three angular frequencies $\omega=10^{-3}$
     (solid lines), $10^{-4}$ (dashed lines) and $10^{-5}$\angfrequnits\ (dash-dotted lines),
     and for the WKB case (dotted lines) as well.
The WKB results are evaluated from equations~(\ref{wkb_eigen1})
     and (\ref{wkb_action_txt}) by using the same value for 
     $\left<u_e^2\right>^{1/2}$ at $R_\odot$ as in the numerical solutions.
It is obvious that for all the frequencies considered, the profiles demonstrate
     substantial deviations from the WKB one, not only in the absolute values 
     but also in spatial dependence.
A WKB-like spatial dependence is recovered only 
     in profiles with $\omega=10^{-3}$\angfrequnits\ and $r\gtrsim 10$~$R_\odot$
     or those with $\omega=10^{-4}$\angfrequnits\ 
     in regions $r\gtrsim 120$~$R_\odot$.

It is interesting to see that, the $\left<b^2\right>^{1/2}$ profiles
     are strikingly similar
     in the region $r\lesssim 10$~$R_\odot$ for all three frequencies.
This behavior can be understood as follows.
First of all, it can be readily shown from equation~(\ref{num_eqs}) that
\begin{eqnarray*}
(M_T^2-1) (|\xi|^2)' = 2\mathrm{Re}(F_{11}) |\xi|^2 +2\mathrm{Re}(F_{12}\eta \xi^*).
\end{eqnarray*}
For the solutions with all three frequencies, 
    it turns out that the first term on the RHS always dominates 
    the second one in the inner corona.
In addition, in the first several solar radii, 
    the flow is very sub-Alfv\'enic,
    i.e., $M_T^2 \ll 1, X_{me} \ll 1$, and $Z\approx -1$.
It follows that $\mathrm{Re}(F_{11})$ is dominated by 
    $-Z(\ln R B_l)'\approx (\ln R B_l)'$.
Therefore $|\xi|^2$ is approximately proportional to
    $1/(R^2 B_l^2)$.
From the definition of $\xi$ (Eq.(\ref{def_xi_eta})), one can see
    that $\left<b^2\right>^{1/2} \sim 1/R$.
That is, $\left<b^2\right>^{1/2}$ shows little frequency dependence. 
On the other hand, from equation~(\ref{wkb_action_txt}) it follows that
    in the WKB limit $\left<b^2\right>^{1/2}\sim \rho^{1/4}$ for $M_T^2\ll 1$ since 
    $U_{ph}\approx U_A$.
Hence the difference between the profiles for the three 
    frequencies and the WKB one in the region $r\lesssim 1.7$\lengthunits\
    reflects the fact that the mass density has a scale height less than
    $R/4$.
Now consider the portion $r \gtrsim 50$\lengthunits\ where $M_T^2 \gg 1$.
One may see that the slope of the $\left<b^2\right>^{1/2}$ profile
    for $\omega=10^{-4}$\angfrequnits\
    becomes similar to that for $\omega= 10^{-3}$\angfrequnits\
    (or equivalently the WKB one) asymptotically.
On the other hand, the profile with $\omega=10^{-5}$\angfrequnits\  
    has a flatter slope than the WKB one.
This is understandable by noting 
    that the $\left<b^2\right>^{1/2}$ profile for $\omega=10^{-5}$\angfrequnits\
    can be rather accurately represented by the zero-frequency solution~(\ref{zero_b}).
By noting that $\rho U_m/B_l$ is a constant
    and $M_T^2 \sim r^{-2}$ for $r \gtrsim 50$~$R_\odot$, one can see that $|\tilde{b}|$ and therefore
    $\left<b^2\right>^{1/2}\sim r^{-1}$.
On the other hand, for the WKB profile one can see from equation~(\ref{wkb_action_txt})
    that asymptotically $\left<b^2\right>^{1/2}\sim \rho^{3/4} \sim r^{-3/2}$
    since $U_{ph} \approx U_m + U_A$. 

Moving on to Figs.\ref{fig_ampltd}b and \ref{fig_ampltd}c, one may notice that
    the magnitude of the ion velocity fluctuations
    $\left<u_k^2\right>^{1/2}$ ($k=p, \alpha$) have nearly the same values at $R_\odot$
    as assumed for $\left<u_e^2\right>^{1/2}$, which can be expected from equation~(\ref{ukandue})
    together with the fact that $\left<b^2\right>^{1/2} \ll B_l$ for $r \approx R_\odot$.
Let us examine Fig.\ref{fig_ampltd}b in some detail.
One can see that between 1~$R_\odot$ and 1~AU,
    the WKB estimate always exceeds the computed $\left<u_p^2\right>^{1/2}$ value.
Furthermore, for $r \lesssim 5$~$R_\odot$,
    the $\left<u_p^2\right>^{1/2}$ profile for
    $\omega=10^{-3}$\angfrequnits\ differs little from that for $\omega=10^{-4}$\angfrequnits,
    however they are significantly larger than that for $\omega=10^{-5}$\angfrequnits.
Take values at 5~$R_\odot$ for instance.
One finds that $\left<u_p^2\right>^{1/2}$ is 83.9\velunits\ for the WKB estimate,
    but 44.6 and 41.9\velunits\ for $\omega=10^{-3}$ and $10^{-4}$\angfrequnits, respectively.
As to the case $\omega=10^{-5}$\angfrequnits, $\left<u_p^2\right>^{1/2}$ is 17.2\velunits.
In other words, the WKB estimate yields a value that is 1.88, 2 and 4.88 times the 
    computed values for $\omega=10^{-3}$, $10^{-4}$, and $10^{-5}$\angfrequnits, respectively.
At 1~AU, the difference between the WKB estimate and computed results is also substantial.
One finds for $\omega=10^{-3}$, $10^{-4}$, and $10^{-5}$\angfrequnits, the WKB estimate
    is 1.83, 1.18 and 1.32 times the values obtained numerically.
An interesting feature of the $\left<u_p^2\right>^{1/2}$ profile for $r\gtrsim 0.3$~AU is that
    it approaches a constant asymptotically.
This can be explained by noting that the magnitudes of ion velocity fluctuations
    for $\omega=10^{-5}$\angfrequnits\ can also be roughly represented 
    by the zero-frequency solution~(\ref{zero_us}).
(The zero frequency solutions~(\ref{zero_exprs})
    can be accurately reproduced if $\omega$ is further reduced.)
At distances $R \gg R_A$ where $M_T^2\gg 1$, equation~(\ref{zero_us}) can be written as
\begin{eqnarray}
\label{zero_up_MTll1}
u_p \approx A_\Omega R \left[\frac{\zeta U_{\alpha p}}{U_p+\zeta U_\alpha}
     -\frac{U_m U_p}{U_j^2}\left(\frac{1}{M_T^2}-\frac{R_A^2}{R^2}\right) \right],
\end{eqnarray}
    where the constant $\zeta = (\rho_\alpha U_\alpha)/(\rho_p U_p)$ 
    is the ion mass flux ratio, and is 0.19
    for the supposed background solution.
In the region considered, the first term is found to dominate
    the second in the square parentheses.
As a result, $u_p\sim R U_{\alpha p} \sim r U_A$.
Since $U_A \sim r^{-1}$, one finds that $|\tilde{u}_p|$ and
    therefore $\left<u_p^2\right>^{1/2}$
    approach a constant asymptotically.

Let us proceed to Fig.\ref{fig_ampltd}c, from which one can see that 
     the $\left<u_\alpha^2\right>^{1/2}$ profiles also deviate considerably
     from the WKB estimate.
In particular, below $\sim 5$~$R_\odot$, the $\left<u_\alpha^2\right>^{1/2}$ profiles 
     behave in a fashion similar to $\left<u_p^2\right>^{1/2}$, however the difference
     between the case for $\omega=10^{-5}$\angfrequnits\ and the WKB estimate
     is even larger.
For instance, at 5~$R_\odot$, the WKB estimate is about 10 times the value
     with $\omega=10^{-5}$\angfrequnits.
On the other hand, for $\omega=10^{-3}$ ($10^{-4}$)\angfrequnits, one finds
     the ratio of the WKB value to that numerically derived
     is 1.89 (2.1).
At larger distances, one may notice the local minimum in the profile
     with $\omega=10^{-3}$\angfrequnits\ at $~0.3$~AU, consistent with Fig.\ref{fig_ReIm}c.
Furthermore, one can see that the values for $\omega=10^{-5}$\angfrequnits\ become
     an order
     of magnitude larger than the WKB one.
At 1~AU, for $\omega=10^{-3}$, $10^{-4}$,
     and $10^{-5}$\angfrequnits, the values obtained numerically
     are 0.55, 1.44 and 11.9 times the WKB one, respectively.
In addition, similar to $\left<u_p^2\right>^{1/2}$,
     $\left<u_\alpha^2\right>^{1/2}$ also approaches a constant value asymptotically.
This is understandable since in the region $R\gg R_A$ where $M_T^2 \gg 1$, one finds
     from equation~(\ref{zero_us}) that
\begin{eqnarray}
\label{zero_ui_MTll1}
u_\alpha \approx A_\Omega R \left[\frac{-U_{\alpha p}}{U_p+\zeta U_\alpha}
     -\frac{U_m U_\alpha}{U_j^2}\left(\frac{1}{M_T^2}-\frac{R_A^2}{R^2}\right) \right],
\end{eqnarray}
     in which the first term in the square parentheses,
     as in the case for $\left<u_p^2\right>^{1/2}$, is found to dominate the second.
Therefore $\left<u_\alpha^2\right>^{1/2} \approx \zeta \left<u_p\right>^{1/2} $ asymptotically.
As a result, $\left<u_\alpha^2\right>^{1/2}$ should show little spatial dependence.

\begin{figure}
\begin{center}
\epsscale{0.9}
\plotone{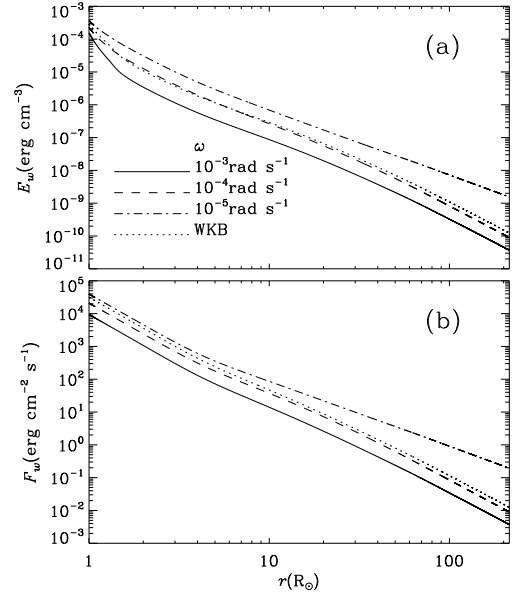}
\caption{
Time-average of (a) the wave energy density $E_w$,
    and (b) wave energy flux density $F_w$ as evaluated from equation~(\ref{def_EFa_wave}).
Numerical results for three angular frequencies as well as for the WKB estimates
    are given in different line styles as indicated in panel (a).
}
\label{fig_EwFw}
\end{center}
\end{figure}

Figure~\ref{fig_EwFw} presents the radial distribution of 
     (a): the wave energy density $E_w$
     and (b): the energy flux density $F_w$
     for the three angular frequencies and WKB expectations.
In both Figs.\ref{fig_EwFw}a and \ref{fig_EwFw}b, the profiles
     for $\omega=10^{-3}$\angfrequnits\ approach a WKB-like behavior when $r\gtrsim 10$~$R_\odot$.
When $\omega=10^{-4}$\angfrequnits, a WKB-like spatial dependence of $E_w$ and $F_w$
     can be seen for $r\gtrsim 80$~$R_\odot$.
Now let us consider Fig.\ref{fig_EwFw}a first.
At 1~$R_\odot$, for
     $\omega=10^{-5}$\angfrequnits, $E_w$ is larger than the WKB result,
     but for $\omega=10^{-3}$ or $10^{-4}$\angfrequnits, it is smaller.
This behavior is determined by the magnetic component in $E_w$
     (Eq.(\ref{def_EFa_wave})) since the particle part is nearly frequency
     independent.
Note that in the WKB limit, the wave energy is equally distributed between 
     the kinetic and magnetic energies,
     $\sum_k \rho_k \left<u_k^2\right>/2 = \left<b^2\right>/(8\pi)$.
Hence at the coronal base, for $\omega=10^{-3}$ and
     $10^{-4}$\angfrequnits, the kinetic energy is larger than the magnetic one,
     whereas for $\omega=10^{-5}$\angfrequnits, the tendency is reversed.
It is interesting to note that among the profiles for the three frequencies
     the highest $\omega$ corresponds to a profile that deviates the most
     from the WKB profile in the region $r\lesssim 2$~$R_\odot$.
In the region $r\gtrsim 30$~$R_\odot$, one can see that
     for $\omega=10^{-4}$\angfrequnits\ the $E_w$ profile
     becomes WKB-like asymptotically.
However, for $\omega=10^{-5}$\angfrequnits\ the slope of the $E_w$ profile
     is flatter than the WKB one.
This can be understood since for $R\gg R_A$, the WKB limit yields
     $E_w \propto |b|^2 \propto \rho^{3/2} \sim r^{-3}$.
On the other hand, for $R\gg R_A$, the zero-frequency result yields
     $E_w$ that is largely determined by the magnetic energy, which leads to
     $E_w \sim |b|^2 \sim r^{-2}$.

Inspection of Fig.\ref{fig_EwFw}b reveals that the
      {the spatial dependence of the profiles of the energy flux density $F_w$ 
      is remarkably similar in the region $r\lesssim 2$~$R_\odot$.
In contrast, in the same region, Fig.\ref{fig_EwFw}a shows that 
      the $E_w$ profiles with different $\omega$ have rather different $r$ dependence.
This behavior of $F_w$} stems from the fact that the
      loss of the wave energy flux
      in the form of the work done on the plasma
      is negligible.
In other words, $F_w$ is diluted only by 
      the flux tube expansion below 2\lengthunits\ (cf. Eq.(\ref{enercons_aver})).
For instance, for $\omega=10^{-4}$\angfrequnits\ at 2~$R_\odot$,
      $\sum_k \rho_k U_k a_{w,k}$ is $43$~erg cm$^{-2}$ s$^{-1}/R_\odot$,
      amounting to only $3.5\%$ of $F_w/r$.
As a result, $(F_w/B_l)' \approx 0$,
     or $F_w \propto B_l$.
{That is, in this region the spatial profile of $F_w$ should show little frequency dependence.}
This is found to be true for all the frequencies considered,
     and for the WKB result as well.
On the other hand, in the region $R\gg R_A$, one can see
     that for $\omega=10^{-4}$\angfrequnits\
     the $F_w$ profile becomes WKB-like asymptotically,
     whereas the profile for $\omega=10^{-5}$\angfrequnits\
     decreases more slowly than the WKB one.
It turns out that asymptotically for the 
     low frequency waves $F_w$ is dominated by the
     Poynting vector which evolves like $\sim \rho U_m R^2 M_T^2 \sim M_T^2 \sim r^{-2}$.
In the WKB limit, however, the ratio of the contributions of 
     particles to the Poynting vector is roughly $1/2$.
As a result, $F_w \sim U_{ph} |b|^2 \sim |b|^2 \sim r^{-3}$. 

\begin{figure}
\begin{center}
\epsscale{.8}
\plotone{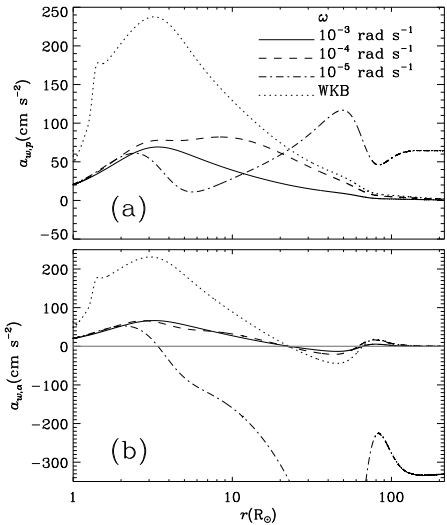}
\caption{
Time-average of the wave acceleration exerted on (a) protons $a_{w, p}$
    and (b) alpha particles $a_{w, \alpha}$ as evaluated from equation~(\ref{def_wave_acce}).
Numerical results for three angular frequencies as well as for the WKB estimates
    are given in different line styles as indicated in panel (a).
In panel (b), the wave acceleration changes sign when the profile crosses the horizontal line.
}
\label{fig_aw}
\end{center}
\end{figure}

From the perspective of solar wind modeling, one may be more curious about the feedback
     from the waves to the background flow.
To this end, Figure~\ref{fig_aw} presents the radial distribution of the acceleration
     exerted on (a) the protons $a_{w,p}$ and (b) alpha particles $a_{w,\alpha}$.
One can see that for $\omega=10^{-3}$ and $10^{-4}$\angfrequnits, the wave acceleration
     is less effective than that in the WKB limit throughout the computational domain.
However, in the low frequency case $\omega=10^{-5}$\angfrequnits, $a_{w,p}$ exceeds
      the WKB expectation considerably beyond 22.8\lengthunits.
Nevertheless, in all cases the wave force tends to accelerate the protons, i.e.,
      $a_{w,p} > 0$ everywhere between the coronal base and 1~AU.
However this is not the case when the alpha particles are concerned.
It can be seen from Fig.\ref{fig_aw}b that the wave acceleration
      $a_{w,\alpha}$ is negative in the interval between 22.7
      and 66.1\lengthunits\ in the WKB limit
      as well in the case $\omega=10^{-3}$\angfrequnits.
This coincidence of the positions where $a_{w,\alpha}$ changes sign
      stems from the fact that beyond $\sim 10$\lengthunits, the wave is WKB-like.
On the other hand, the $a_{w, \alpha}$ profile for $\omega=10^{-4}$\angfrequnits\
      changes sign at 21.8 and 61.8\lengthunits.
{These locations are slightly different from their counterparts in the WKB limit.}
When it comes to $\omega=10^{-5}$\angfrequnits, the $a_{w,\alpha}$ profile does not show  
      any resemblance to the WKB expectation.
In particular, the waves start to decelerate the alpha particles even in the inner 
      corona: $a_{w, \alpha}$ is negative everywhere beyond $3.4$\lengthunits.
An interesting aspect of the wave-induced acceleration is that asymptotically 
      both $a_{w,p}$ and $a_{w,\alpha}$ approach constant values.
This is understandable since in the zero-frequency limit, 
      in the region where $R \gg R_A$, both $u_p$ and $u_k$ show little
      radial dependence.
It then follows from the expressions~(\ref{zero_up_MTll1}) and (\ref{zero_ui_MTll1})
      that for ion species $k$ ($k=p, \alpha$), 
$a_{w,k} \sim u_k^2 (\ln R)' - U_k  u_k \left(\ln R\right)' b/B_l 
        \propto R' u_k$.
Since in the region considered, the line of force is nearly perfectly radial, one can see
      that $R'$ is a constant. 
As a result, $a_{w,k}$ should approach a constant asymptotically.

\begin{figure}
\begin{center}
\epsscale{1.1}
\plotone{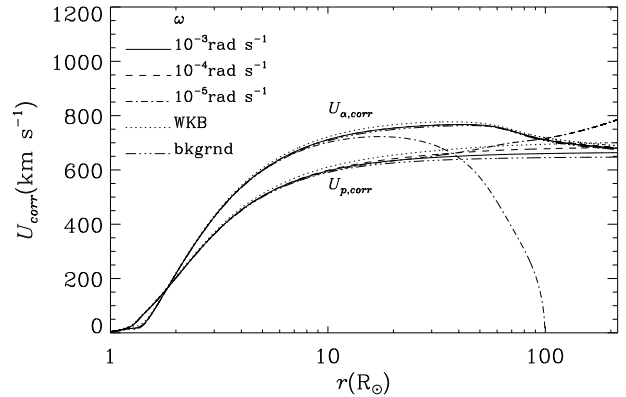}
\caption{
Possible modification of ion flow speeds due to the wave acceleration.
Although the speeds $U_{p, corr}$ and $U_{\alpha, corr}$, evaluated from
     equation~(\ref{def_ucorr}), 
     are not computed self-consistently,
     their deviation from the background flow parameters represents the significance
     of the wave acceleration in the force balance for the two ion species.
Numerical results for three angular frequencies as well as for the WKB estimates
    are given in different line styles as indicated.
}
\label{fig_ucorr}
\end{center} 
\end{figure}

The extent to which the wave forces may alter the ion flows can be obtained only 
     through a self-consistent modeling by using, e.g., the iterative
     approach adopted by \cite{MacChar_94}:
     The wave equation~(\ref{num_eqs}) and the solar wind
     equations~(\ref{bkg_nk}) to (\ref{bkg_ps}) incorporating
     the wave contribution
     are solved alternately until a convergence is met.
As a first step, however, we may simply evaluate the ion
     speeds $U_{k, corr}$ ($k=p, \alpha$) corrected for
     the wave force, i.e.,
\begin{eqnarray}
U_{k, corr}^2 - U_k^2= 2 \int_{R_\odot}^r a_{w,k} \mathrm{d}l. 
\label{def_ucorr}
\end{eqnarray}
Figure~\ref{fig_ucorr} presents the radial distribution of both
      $U_{p, corr}$ and $U_{\alpha, corr}$ for all the frequencies
      considered.
The background flow speed profiles are also plotted for comparison.
One may see that, with the present choice of the wave amplitude,
      the waves have negligible effects on the ion acceleration
      below the Alfv\'en point.
Beyond the Alfv\'en point, the effects introduced by the waves on
      the speed profiles become more important, especially in
      the low frequency case.
As a matter of fact, for $\omega=10^{-5}$\angfrequnits\ the corrected proton speed $U_{p, corr}$
      reaches 785\velunits\ at 1~AU where the background value is 648\velunits.
As for the alpha speed, $U_{\alpha, corr}$ becomes negative beyond 
      100\lengthunits\ due to the significant deceleration
      exerted on the alphas by the waves.
Of course, such a situation will not appear in reality. 
What will happen is that the protons are accelerated whereas alphas are decelerated
      by the low frequency waves until the ions move at nearly identical speeds.
The net effect of low-frequency waves is thus to limit the speed difference between 
      the protons and alpha particles at large distances.
However, it should be pointed out that in these regions
      the net work done by the low-frequency wave
      on the solar wind as a whole is negligible: $\sum_k\rho_k U_k a_{w,k} \approx 0$
      and $F_w$ is nearly divergence free
      as discussed in reference to Fig.\ref{fig_EwFw}.

\section{SUMMARY AND CONCLUDING REMARKS}
\label{sec_conc}
This study has been motivated by the apparent lack of a non-WKB analysis of Alfv\'en waves
      in a multi-fluid solar wind with differentially flowing ions.
To be more specific, this study is concerned with the propagation of
      dissipationless, hydromagnetic (angular frequency $\omega$ well below ion gyro-frequencies),
      purely toroidal Alfv\'en waves 
      that propagate in a background 3-fluid solar wind comprised of
      electrons, protons and alpha particles.
Azimuthal symmetry is assumed throughout.
No assumption has been made that the wavelength is small compared with
      the spatial scales at which
      the background flow parameters vary.
The wave behavior at a given $\omega$ is governed by equation~(\ref{num_eqs}), which
      is derived from the general transport equations in the five-moment approximation.
The Alfv\'enic point, where the combined Alfv\'en Mach number $M_T=1$
      (cf. Eq.(\ref{def_mt2})), is a singular point of equation~(\ref{num_eqs})
      and a regularity condition has to be imposed.
For the other boundary condition, we impose a velocity amplitude
      of 10\velunits\ at the coronal base (1~$R_\odot$).
For the given background model of a realistic low-latitude fast solar wind, 
      equation~(\ref{num_eqs}) is integrated numerically for
      three representative angular frequencies
      $\omega=10^{-3}$, $10^{-4}$ and $10^{-5}$\angfrequnits\ to yield the radial distribution 
      of the wave energy and energy flux densities as well as the wave-induced acceleration
      exerted on ion species.

The first conclusion concerns the applicability of the WKB approximation.
Between 1~$R_\odot$ and 1~AU, the
      numerical solutions show substantial deviation from the WKB expectations.
Even for the relatively high frequency $\omega=10^{-3}$\angfrequnits,
      a WKB-like behavior can be seen only in regions where $r\gtrsim 10$~$R_\odot$.
In the low-frequency case $\omega=10^{-5}$\angfrequnits, the computed profiles
      of wave-related parameters show a spatial dependence that is
      distinct from the WKB one, the deviation being particularly pronounced
      in interplanetary space.
In the inner corona $r\lesssim 4$~$R_\odot$,
      the computed ion velocity fluctuations are
      considerably smaller than the WKB expectations in all cases,
      as is the computed wave-induced acceleration exerted on 
      protons or alpha particles.
As for the wave energy and energy flux densities, they can be enhanced
      or depleted compared
      with the WKB results, depending on $\omega$.

The second conclusion is concerned with how the wave acceleration may alter the background
      flow parameters.
In reference to Fig.\ref{fig_ucorr}, it is found that with the current choice of
      base wave amplitude, the wave acceleration has little effect on the force balance
      for protons or alpha particles in the corona.
That is, one has to invoke processes other than the non-WKB wave acceleration to accelerate
      the ions out of the gravitational potential well of the Sun.
However, at large distances beyond the Alfv\'enic point, low-frequency waves may play an important role
      in the ion dynamics, with the net effect being to equalize the speeds of 
      the two ion species considered.

Strictly speaking, the separation of the flow into fluctuations and a time-independent background
      implies that the waves are linear.
However, one may have noticed that the wave amplitude $\left<b^2\right>^{1/2}$ at 1~AU for
      $\omega=10^{-5}$~s$^{-1}$\angfrequnits\ is substantially larger than the background
      poloidal magnetic field strength (cf. Fig.\ref{fig_ampltd}a).
That the transverse magnetic field dominates the poloidal one 
      demands a careful examination of the nonlinear effects other than the wave-induced
      acceleration.
In particular, one needs to look at the generation of secondary waves
      and structures by the primary Alfv\'en waves through the source terms
      in the momentum equation.
As discussed by \citet{Lou_93}, in the case of ideal MHD these source terms decrease 
      sufficiently fast with radial distance asymptotically.
Consequently, the first-order wave amplitudes
      are valid provided that the amplitude imposed at the coronal base
      is sufficiently small.
The basic picture is expected to be the same even if a second ion species is included,
      although a similar discussion in the 3-fluid framework will be complicated by the 
      richness of wave modes due to the differential proton-alpha streaming
      \citep[e.g.,][]{McKenzie_etal_93}.

As has been mentioned in the introduction, Alfv\'en waves
      are dissipated in some way, and this dissipation of the primary waves
      should be described self-consistently.
A possibility to do this is to perform a full \elsasser\ analysis
      extended to the multi-fluid case, and to express the dissipation
      in terms of the amplitudes of counter-propagating waves.
We note that this already complicated issue will become even trickier considering 
      the necessity to apportion the dissipated wave energy among different species.

In closing, we note that the low-frequency waves may also be important for outflows from
     stars other than the Sun.
For instance, in the radiatively driven stellar winds, these waves will provide
     a further channel of momentum exchange between passive ions and line-absorbing ions
     in addition to the Coulomb friction.
This possibility was first pointed out by \citet{Pizzo_etal_83} in connection with the effects
     of stellar rotation.
Due to the clear resemblance between the low-frequency Alfv\'en waves and stellar rotation
     (cf. section~\ref{sec_zero_freq}), 
     their discussion also applies to the case where the star persistently
     emits Alfv\'en waves with frequencies lower than the critical one
     defined by equation~(\ref{def_omegac}).
Consequently, the mass loss rate may be significantly altered.
A quantitative study of this effect is beyond the scope of the present paper though.

\acknowledgements
This research is supported by a PPARC rolling grant to the University of Wales Aberystwyth.
We thank Shadia Rifai Habbal for her careful reading of the first draft,
        and Hui-Nan Zheng for helpful discussions.

% 
% 

% 

% 
% 

% 
% 

% 
% \clearpage
 

\begin{thebibliography}{}
\bibitem[Alazraki \& Couturier(1971)]{AlazrakiCouturier_71}
Alazraki, G., \& Couturier, P. 1971, {\aap}, 13, 380

\bibitem[Armstrong \& Woo(1981)]{ArmstrongWoo_81}
Armstrong, J.~W., \& Woo, R. 1981, {\aap}, 103, 415

\bibitem[Banaszkiewicz et al.(1998)]{Bana_etal_98}
Banaszkiewicz, M., Axford, W.~I., \& McKenzie, J.~F. 1998, 
{\aap}, 337, 940

\bibitem[Banerjee et al.(1998)]{Banerjee_etal_98}
Banerjee, D., Teriaca, L., Doyle, J.~G., \& Wilhelm, K. 1998, {\aap}, 339, 208

\bibitem[Bavassano et al.(2000a)]{Bavassano_etal_00a}
Bavassano, B., Pietropaolo, E., \& Bruno, R. 2000a, {\jgr}, 105, 12697

\bibitem[Bavassano et al.(2000b)]{Bavassano_etal_00b}
Bavassano, B., Pietropaolo, E., \& Bruno, R. 2000b, {\jgr}, 105, 15959

\bibitem[Belcher \& Davis(1971)]{BelcherDavis_71}
Belcher, J.~W., \& L. Davis, Jr. 1971, {\jgr}, 76, 3534


\bibitem[Cranmer \& van Ballegooijen(2005)]{CranmerBalle_05}
Cranmer, S.~R., \& van Ballegooijen, A.~A. 2005, {\apjs}, 156, 265


\bibitem[Dmitruk et al.(2001)]{Dmitruk_etal_01}
Dmitruk, P., Milano, L.~J., \& Matthaeus, W.~H. 2001, {\apj}, 548, 482


\bibitem[Esser et al.(1999)]{Esser_etal_99}
Esser, R., Fineschi, S., Dobrzycka, D., et al. 1999, {\apj}, 510, L63


\bibitem[Goldstein et al.(1995)]{Goldstein_etal_95}
Goldstein, M.~L., Roberts, D.~A., \& Matthaeus, W.~H. 1995, {\araa}, 33, 283


\bibitem[Heinemann \& Olbert(1980)]{HO80} 
Heinemann, M. \& Olbert, S. 1980, {\jgr}, 85, 1311


\bibitem[Hollweg(1974)]{Hollweg_74}
Hollweg, J.~V. 1974, {\jgr}, 79, 1357

\bibitem[Hollweg et al.(1982)]{Hollweg_etal_82}
Hollweg, J.~V., Bird, M.~K., Volland, H., et al. 1982, 
     {\jgr}, 87, 1

\bibitem[Hollweg \& Isenberg(2002)]{HI02}
Hollweg, J.~V., \& Isenberg, P.~A. 2002, {\jgr}, 107(A7), doi:10.1029/2001JA000270


\bibitem[Isenberg \& Hollweg(1982)]{IH82}
Isenberg, P.~A., \& Hollweg, J.~V. 1982, {\jgr}, 87, 5023

\bibitem[Li et al.(1999)]{Li_etal_99}
Li, X., Habbal, S.~R., Hollweg, J.~V., \& Esser, R. 1999, {\jgr}, 104, 2521


\bibitem[Li \& Li(2006)]{LiLi06}
Li, B., \& Li, X. 2006, {\aap}, 456, 359


\bibitem[Lou(1993)]{Lou_93}
Lou, Y.-Q. 1993, {\jgr}, 98, 3563

\bibitem[MacGregor \& Charbonneau(1994)]{MacChar_94}
MacGregor, K.~B., \& Charbonneau, P. 1994, {\apj}, 430, 387


\bibitem[McComas et al.(2000)]{McComas_etal_00}
McComas, D.~J., Barraclough, B.~L., Funsten, H.~O., et al. 2000, {\jgr}, 105, 10419

\bibitem[McKenzie et al.(1979)]{McKenzie_etal_79}
McKenzie, J.~F., Ip, W.-H., \& Axford, W.~I. 1979, {\apss}, 64, 183

\bibitem[McKenzie et al.(1993)]{McKenzie_etal_93}
McKenzie, J.~F., Marsch, E., Baumg\"{a}rtel, K., \& Sauer, K. 1993, 
{Ann. Geophysicae}, 11, 341

\bibitem[McKenzie(1994)]{McKenzie_94}
McKenzie, J.~F. 1994, {\jgr}, 99, 4193


\bibitem[Marsch et al.(1982)]{Marsch_etal_82}
{Marsch, E., M\"{u}hlh\"{a}user, K.-H., Rosenbauer, K., Schwenn, R., \& Neubauer, F.~M.} 1982, 
{\jgr}, {87}, 35

\bibitem[Parker(1965)]{Parker_65}
Parker, E.~N. 1965, {\ssr}, 4, 666


\bibitem[Pizzo et al.(1983)]{Pizzo_etal_83}
Pizzo, V., Schwenn, R., Marsch, E. et al. 1983, {\apj}, 271, 335


\bibitem[Scott et al.(1983)]{Scott_etal_83}
Scott, S.~L., Coles, W.~A., \& Bourgois, G. 1983, {\aap}, 123, 207

\bibitem[Smith \& Balogh(1995)]{SmithBalogh_95}
Smith, E.~J., \& Balogh, A. 1995, {\grl}, 22, 3317 

\bibitem[Tu \& Marsch(1995)]{TuMarsch_95}
Tu, C.-Y., \& Marsch, E. 1995,  
{\ssr}, 73, 1

\bibitem[Verdini et al.(2005)]{Verdini_etal_05}
Verdini, A., Velli, M., \& Oughton, S. 2005, 
{\aap}, 444, 233
\end{thebibliography}
\end{document}